\documentclass[conference]{IEEEtran}
\IEEEoverridecommandlockouts
\usepackage{amsmath,amssymb,amsfonts}
\usepackage{algorithmic}
\usepackage{graphicx}
\usepackage{textcomp}
\usepackage{xcolor}
\usepackage{cite}
\usepackage{balance}
\usepackage{microtype}
\usepackage{url}

\def\BibTeX{{\rm B\kern-.05em{\sc i\kern-.025em b}\kern-.08em
    T\kern-.1667em\lower.7ex\hbox{E}\kern-.125emX}}

\usepackage{booktabs}
\usepackage{caption}
\usepackage{tabularx}
\usepackage{multirow}
\newcommand{\mytable}{
    \centering
    \small
    \renewcommand{\arraystretch}{1.2}
    }

\newcolumntype{C}{>{\centering\arraybackslash}X}
\newcolumntype{L}{>{\raggedright\arraybackslash}X}

\usepackage{amsfonts}
\usepackage{bm}

\usepackage[prependcaption,textsize=scriptsize]{todonotes}
\definecolor{mycolor}{HTML}{FF6600}

\begin{document}

\title{
Unsupervised Word Discovery: Boundary Detection with Clustering vs. Dynamic Programming
\thanks{This work was supported through a grant from Fab Inc.}%
}

\author{
    % \and
    \IEEEauthorblockN{
        Simon Malan, Benjamin van Niekerk, Herman Kamper
    }
    \IEEEauthorblockA{
        \textit{Electrical and Electronic Engineering,         Stellenbosch University, South Africa}\\
        {
            % \footnotesize \tt
            \small
            24227013@sun.ac.za, benjamin.l.van.niekerk@gmail.com, kamperh@sun.ac.za
        }
    }
    % \and
}

% \author{\IEEEauthorblockN{Simon Malan}
% \IEEEauthorblockA{\textit{Electrical and Electronic Engineering} \\
% \textit{Stellenbosch University}\\
% 24227013@sun.ac.za}
% \and
% \IEEEauthorblockN{Benjamin van Niekerk}
% \IEEEauthorblockA{\textit{Electrical and Electronic Engineering} \\
% \textit{Stellenbosch University}\\
% benjamin.l.van.niekerk@gmail.com }
% \and
% \IEEEauthorblockN{Herman Kamper}
% \IEEEauthorblockA{\textit{Electrical and Electronic Engineering} \\
% \textit{Stellenbosch University}\\
% kamperh@sun.ac.za}
% }

\maketitle

\begin{abstract}
We look at the long-standing problem of segmenting unlabeled speech into word-like segments and clustering these into a lexicon. Several previous methods use a scoring model coupled with dynamic programming to find an optimal segmentation. Here we propose a much simpler strategy: we predict word boundaries using the dissimilarity between adjacent self-supervised features, then we cluster the predicted segments to construct a lexicon. For a fair comparison, we update the older ES-KMeans dynamic programming method with better features and boundary constraints. On the five-language ZeroSpeech benchmarks, our simple approach gives similar state-of-the-art results compared to the new ES-KMeans+ method, while being almost five times faster.
Project webpage: \url{https://s-malan.github.io/prom-seg-clus}.
\end{abstract}

\begin{IEEEkeywords}
word segmentation, lexicon learning, zero resource speech processing, unsupervised learning
\end{IEEEkeywords}

\section{Introduction}
\label{sec:intro}

Unsupervised word segmentation aims to identify word-like segments in raw speech audio.
This is challenging since speech is a continuous stream 
without obvious silences between words~\cite{okko_early_lan_acq}. 
Constructing a lexicon poses another challenge, as no two speakers are identical and
even individual speakers show a lot of variation in their speech.
Remarkably, human infants navigate these challenges, demonstrating
word discrimination and recognition capabilities within their first year~\cite{elika_6_9_word_meaning, native_lan_acq}. 
Solving the problem of segmenting and clustering speech could provide a way to improve
our understanding of human language acquisition~\cite{emmanuel_la_reverse_eng}.
It could also advance the development of low-resource speech technologies~\cite{aaurent_asr_low_resource}.

Early word discovery methods
employed direct pattern matching, usually using dynamic time-warping~\cite{park_seg_dtw}, to find matching segments in pairs of utterances. 
Although these methods have shown
progress~\cite{dusted}, they fail to discover patterns that cover all the speech audio. 
In this paper we are particularly interested in full-coverage systems that provide a full tokenization of the input speech into word-like units.

Several full-coverage methodologies have been considered~\cite{lee+etal_tacl15,algayres+etal_tacl22,okuda2022double}.
One strand of methods discovers phone-like units and does word segmentation and lexicon learning 
on top of (or in conjunction with)
the subword units~\cite{jansen+etal_icassp13,bhati_scpc,cuervo2022contrastive}.
The recent duration-penalized dynamic programming (DPDP) method is one example~\cite{herman_dpdp, herman_dpdp_hu}. 
It uses quantized self-supervised speech representations for subword learning and an autoencoding recurrent model for subsequent word learning.
Another strand of methods models higher-level units like words directly without explicit subword modeling~\cite{okko_sylseg,herman_bes_gmm,wang+etal_icassp18,fuchs2023unsupervised}.
The older embedded segmental $K$-means (ES-KMeans) method is an example~\cite{herman_eskmeans}.
It uses an iterative segmenting and clustering scheme, each time choosing the current best boundary hypothesis using dynamic programming (similar to DPDP).
Surprisingly, ES-KMeans is still competitive~\cite{herman_dpdp_hu}, despite using older speech features and boundary constraints.

In this paper we propose a much simpler full-coverage speech segmentation system that does not require dynamic programming.
Inspired by Pasad et al.~\cite{ankita_tti}, we argue that (1)~well-defined word boundaries can be found through a lightweight method measuring dissimilarity between adjacent self-supervised features, and (2)~an explicit lexicon can be constructed by just doing $K$-means on the discovered word segments (if appropriate segmental features are used).
We compare this method to several other approaches -- including our own updated version of ES-KMeans -- on English, French, Mandarin, German, and Wolof evaluations from Track 2 of the ZeroSpeech Challenge~\cite{ewac_zrc}.

We make the following contributions.
(1)~We show that the combination of good word boundaries with high-quality self-supervised segmental features can compete with dynamic programming methods while being much faster.
(2)~We introduce ES-KMeans+, an updated version of ES-KMeans that is efficient and achieves some of the best results on the ZeroSpeech benchmarks.
(3)~We investigate how the
pre-training languages in the self-supervised speech models affects our simple method. We find that language-specific models trained on the target language outperform multilingual models.

\section{Prominence-Based Boundaries with Clustering}
\label{sec:method}

Our simple full-coverage unsupervised word segmentation system consists of two components. 
First, we determine word boundaries using a simple prominence-based approach. 
Second,
we cluster these predicted word-like units to build a lexicon.

\begin{figure}[t]
    \centerline{\includegraphics[width=0.99\columnwidth]{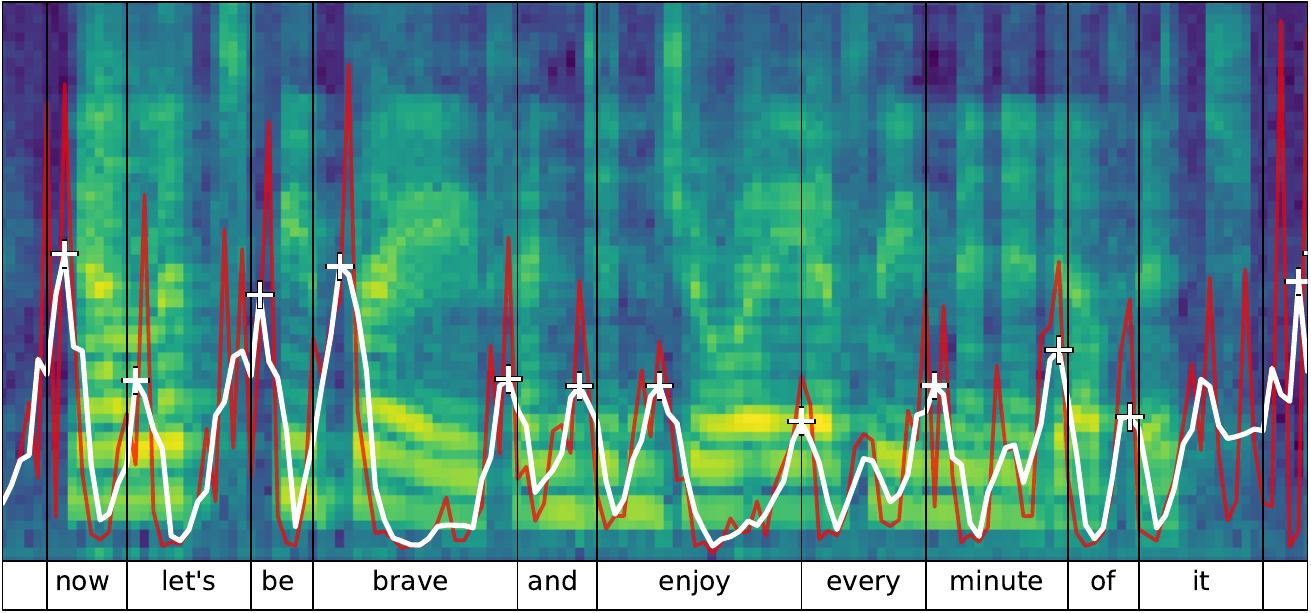}}
    \caption{An example of word boundaries from the prominence-based approach of~\cite{ankita_tti}. The red (dark) line is the dissimilarity curve between adjacent frames, which is smoothed to produce the white line. The crosses are the predicted boundaries. The black vertical lines are the ground truth boundaries.}
    \label{fig:tti}
\end{figure}

For word boundary detection, we follow the method of Pasad et al.~\cite{ankita_tti}.
This lightweight method extracts high-quality word boundaries without the need for training an explicit boundary detection model.
Speech utterances are first encoded by extracting features from an intermediate HuBERT~\cite{wei_hubert} layer, which we denote as $\mathbf{y}_{1:T}=\mathbf{y}_{1},\mathbf{y}_{2},...,\mathbf{y}_{T}$.
To predict word boundaries, the dissimilarity between neighboring frames is calculated using the cosine distance
$f_{t}=d(\mathbf{y}_{t+1},\mathbf{y}_{t})$
between adjacent frames. 
Hereafter, a smoothing function, using a moving average, is applied to the dissimilarity curve.
(For this step,
the HuBERT features are
mean and variance normalized beforehand.)

Peaks on the smoothed dissimilarity curve, as seen in Fig.~\ref{fig:tti}, are marked as word boundaries when the dissimilarity at frame $t$ is greater than some prominence threshold. 
Using prominence instead of a hard dissimilarity threshold provides more nuanced boundaries since the prominence of the curve indicates how dissimilar the current frames are relative to the dissimilarity of the surrounding frames.
The
method works on the principle that frames within the same word are close to each other, while frames at word boundaries are further away from each other. 
The features are crucial as
HuBERT focuses on encoding phonetic information while throwing away speaker-specific information~\cite{wei_hubert, bshall_s_vs_d_units}.

\begin{figure}[t]
    \centerline{\includegraphics[width=0.99\columnwidth]{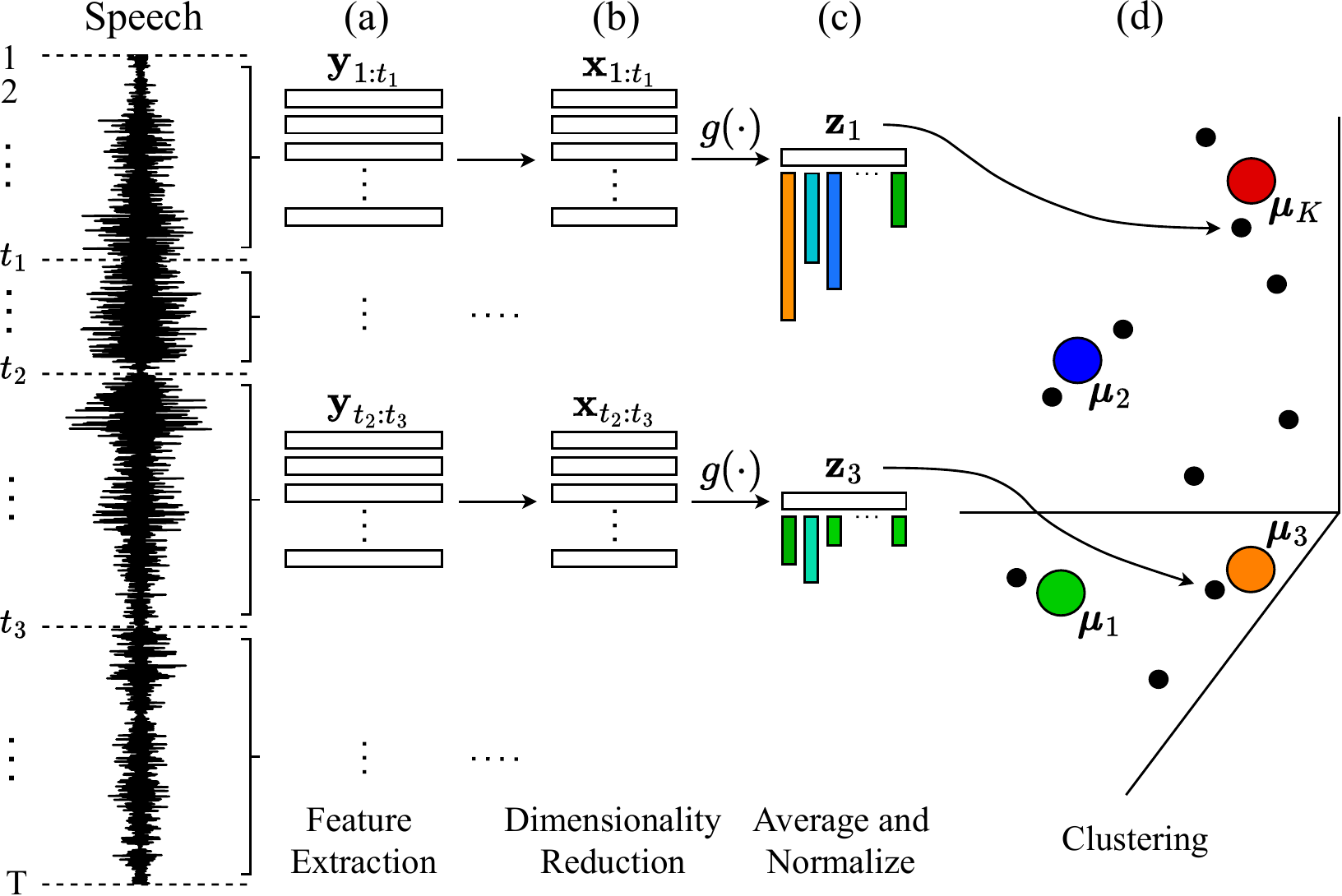}}
    \caption{
    Our lexicon building step. After extracting frame-level features (a), PCA dimensionality reduction is applied (b). For each segment from the prominence-based approach (Fig.~\ref{fig:tti}), an averaged embedding is obtained~(c). These are $K$-means clustered (d) to get a lexicon.}
    \label{fig:cluster}
\end{figure}

The lexicon building step, illustrated in Fig.~\ref{fig:cluster}, takes the word boundaries (the dashed lines on the left side of the figure) and their corresponding speech utterances as input.
Again, utterances are encoded with an intermediate HuBERT layer, resulting in speech features in a high dimensional space $\mathbf{y}_{t}\in~\mathbb{R}^{D}$ (a in the figure).
For clustering, working with these high-dimensional features can become
computationally expensive. 
Therefore, we apply PCA dimensionality reduction (b) to the features, reducing them to a lower dimensional space
$\mathbf{x}_{t}\in \mathbb{R}^{M}$, with $M < D$,
without loosing much phonetic information~\cite{ramon_awe}.
In this case, the HuBERT features are not normalized and can come from a different layer than the one used for boundary detection, i.e.\ we overload the symbol $\mathbf{y}$.

To cluster the variable duration word segments, we turn to acoustic word embeddings, which map variable-length speech segments to fixed dimensional vectors~\cite{andrew_awe, keith_awe, kamper_awe}.
We specifically follow the simple
approach of~\cite{ramon_awe}, where an embedding is obtained by averaging the features in the predicted word segment. 
Concretely, a word segment $\mathbf{x}_{t_{1}:t_{2}}$ is transformed into a fixed dimensional embedding vector $\mathbf{z}_{i}=g(\mathbf{x}_{t_{1}:t_{2}})$, where $g$ represents averaging followed by normalization to the unit sphere. 
The result is a set of embeddings $\mathbf{z}_{i} \in \mathbb{R}^M$ (c in Fig.~\ref{fig:cluster}) that are clustered using $K$-means clustering (d).
Our implementation uses the efficient
FAISS\footnote{\url{https://github.com/facebookresearch/faiss}} library for clustering. 
As illustrated on the right side of the figure, each word segment is assigned to a class $k$ whose centroid, $\bm{\mu}_{k}$, is closest to the segment embedding.

\begin{table*}[!t]
    \mytable
    \captionsetup{justification=centering}
    \caption{Performance (\%) of prominence segmentation with clustering on English HuBERT, and other state-of-the-art methods for word segmentation and lexicon building on Track 2 of the ZeroSpeech Challenge.}
    \begin{tabularx}{\linewidth}{@{}l@{\ \ }C@{\ \ }C@{\ \ }CC@{\ \ }C@{\ \ }CC@{\ \ }C@{\ \ }CC@{\ \ }C@{\ \ }CC@{\ \ }C@{\ \ }C@{}}    
        \toprule
        & \multicolumn{3}{c}{English} & \multicolumn{3}{c}{French} & \multicolumn{3}{c}{Mandarin} & \multicolumn{3}{c}{German} & \multicolumn{3}{c}{Wolof}\\
        \cmidrule{2-4}
        \cmidrule(l){5-7}
        \cmidrule(l){8-10}
        \cmidrule(l){11-13}
        \cmidrule(l){14-16}
        Model & NED & $R$-val. & Token $F_{1}$ & NED & $R$-val. & Token $F_{1}$ & NED & $R$-val. & Token $F_{1}$ & NED & $R$-val. & Token $F_{1}$ & NED & $R$-val. & Token $F_{1}$\\
        \midrule
        VG-HuBERT~\cite{vg_hubert} & 41.0 & 59.8 & \textbf{24.0} & 62.0 & 44.0 & 15.0 & 73.0 & 32.5 & 19.0 & 56.0 & 21.9 & \textbf{15.0} & 92.0 & 59.7 & 9.0 \\
        DPDP~\cite{herman_dpdp_hu} & 41.7 & \textbf{63.2} & 19.6 & 66.0 & 60.3 & 11.6 & 86.0 & \textbf{67.9} & 24.5 & 56.8 & \textbf{49.2} & 12.5 & 72.2 & 66.9 & 13.1 \\ 
        ES-KMeans~\cite{herman_eskmeans} & 73.2 & 51.5 & 19.2 & 68.7 & 37.2 & 6.3 & 88.1 & 23.3 & 8.1 & 66.2 & 17.1 & 11.5 & 72.4 & 58.3 & 10.9 \\
        ES-KMeans+~[ours]& 33.5 & 50.0 & 14.7 & \textbf{43.2} & 56.3 & \textbf{20.0} & \textbf{65.5} & 56.9 & \textbf{25.1} & \textbf{42.8} & 25.5 & 9.7 & \textbf{56.2} & \textbf{69.4} & \textbf{24.5} \\
        Prom. Seg. Clus.~[ours] & \textbf{32.9} & 60.6 & \textbf{24.0} & 47.9 & \textbf{61.0} & 17.2 & 71.4 & 58.2 & 22.7 & 44.3 & 41.8 & 10.9 & 59.3 & 67.9 & 19.5 \\
        \bottomrule
    \end{tabularx}
    \label{tbl:zrc_results}
\end{table*}

\section{Dynamic Programming Methods}
\label{sec:dp_methods}

We will compare our approach to state-of-the-art dynamic progamming based methods.
Duration-penalized dynamic programming (DPDP) is a two-stage method that first tokenizes input speech into phone-like units and then segments the phone tokens into word-like units~\cite{herman_dpdp}.
The original
method did word segmentation without lexicon learning, but~\cite{herman_dpdp_hu} extended DPDP using $K$-means on averaged HuBERT features to get a lexicon.

In~\cite{herman_dpdp} and~\cite{herman_dpdp_hu}, DPDP is compared to the much older embedded segmental $K$-means (ES-KMeans) method~\cite{herman_eskmeans}.
In ES-KMeans, potential word boundaries are iteratively clustered and re-segmented using dynamic programming until the best boundaries are selected to form the final word-like segments. 
To constrain the huge number of possible boundary positions, potential word boundary positions
are determined by SylSeg~\cite{okko_sylseg}, an unsupervised syllabification method using signal processing techniques.
The original ES-KMeans uses
subsampled MFCC features to represent the word-like segments.
Although DPDP outperforms ES-KMeans,
this comparison is unfair due to the outdated nature of ES-KMeans. 
To level the playing field, we improve the components of ES-KMeans and call our updated approach ES-KMeans+. 

Concretely, we replace the SylSeg boundary constraints with the prominence-based approach described in Section~\ref{sec:method}. 
We replace MFCCs with HuBERT features and follow the same clustering approach as in Fig.~\ref{fig:cluster} (PCA projected segment features are averaged, normalized, and then iteratively clustered).
We change the original ES-KMeans algorithm which iterated over individual utterances to instead operate over batches. This allows us to again use the efficient FAISS library for clustering.
The benefits of each of these changes were verified in developmental experiments, leading to a much improved and scalable implementation.

After these updates, ES-KMeans+ and the prominence-based approach of Section~\ref{sec:method} are very similar. The latter corresponds to the first iteration of ES-KMeans+, just before word segmentation is performed. The prominence-based approach is therefore much faster, but ES-KMeans+ has the benefit that it can decide to remove prominence-detected boundaries if these are poorly matched to the clustering model.

\section{Experimental Setup}
\label{sec:exp_setup}

We perform our main experiments on Track 2 of the ZeroSpeech Challenge~\cite{ewac_zrc}. 
This covers
five languages: English, French, Mandarin, German, and Wolof, respectively consisting of 45, 24, 2.5, 25, and 10 hours of speech. 
The challenge encourages participants to develop language-invariant methods and is one of the most comprehensive benchmarks for unsupervised speech systems.
For development,
we use the dev-clean subset of LibriSpeech~\cite{librispeech}
with 5.4 hours of English speech from 40 speakers.

Word segmentation is evaluated using $R$-value and token $F_{1}$, where higher is better for both.
$R$-value measures how close hypothesized word boundaries are to an 
ideal operating point with a
100\% hit-rate and 0\% over-segmentation~\cite{okko_rval}. 
Token $F_{1}$ evaluates how well the hypothesized word tokens match ground truth tokens, requiring both predicted boundaries to be correct in order to receive credit.

The quality of a lexicon is evaluated using normalized edit distance (NED)~\cite{ned}, which relies on phonemic transcriptions found by forced alignments. 
Discovered word tokens are mapped to their overlapping phoneme sequence, and the NED is calculated between all phoneme sequences of the segments within each cluster. Lower NED is better.

For both the prominence-based word segmentation step of Section~\ref{sec:method} and for the potential boundary set of ES-KMeans+ in Section~\ref{sec:dp_methods}, we extract features from the 9th HuBERT layer, based on experiments in~\cite{ankita_tti}. 
The smoothing window controls how emphasized the dissimilarity curve is, while the prominence threshold determines which peaks are chosen as word boundaries.
We set these parameters
by finding a 
balance between 
NED and $R$-value on the development data.
For our prominence-based approach, we select a four-frame window 
with a 0.75 prominence threshold, while for ES-KMeans+ 
we opt for a five-frame window with a threshold of 0.3. 
The high-recall hyperparameter setting for ES-KMeans+ creates
more word boundaries than are needed, enabling the method to choose the best subset of these boundaries.

When clustering, both in our prominence-based method and ES-KMeans+, we extract features from the 12th HuBERT layer and reduce the feature dimensionality to 250 dimensions using PCA.
These settings are based on development experiments.
To enable a fair comparison to previous work,
we use the same number of $K$-means clusters as in the original ES-KMeans and DPDP papers: we use 43k, 29k, 3k, 29k, and 3.5k clusters for English, French, Mandarin, German, and Wolof.

We compare our system to the dynamic programming methods of Section~\ref{sec:dp_methods}: DPDP, ES-KMeans, and ES-KMeans+. 
We also compare to the visually-grounded HuBERT (VG-HuBERT) method~\cite{vg_hubert} that pairs unlabeled speech with images.

\section{Experimental Results}
\label{sec:results}

We start our experiments with a comparison to previous full-coverage unsupervised segmentation and clustering systems, with a specific focus on comparing our new simple approach to the previous dynamic programming methods that inspired it.
We then look at the impact of different design choices within our simple method, and finally consider
the impact of using an English self-supervised speech model on non-English data.

\subsection{Comparison to Other Systems}

The performance of all systems can be seen in Table~\ref{tbl:zrc_results}.
When comparing 
the prominence-based method to 
ES-KMeans+,
we see
a tradeoff between NED and $R$-value: the simple method achieves better $R$-value on all languages except Wolof, while ES-KMeans+ gives better NED in all cases but English.
Compared to the other approaches, 
both our systems achieve several 
state-of-the-art results and improve upon ES-KMeans. 
Prominence segmentation achieves a good tradeoff between metrics, finding a 
middle ground between ES-KMeans+ and DPDP (which 
consistently achieves a high $R$-value).

One
limitation of the prominence-based method is its 
dependence on the quality of the word segmentation step 
since it cannot remove bad boundaries like DPDP or ES-KMeans can.
But, while
ES-KMeans+ is marginally better than the prominence-based 
approach on several metrics, Table~\ref{tbl:runtime} shows that ES-KMeans+ is four to five times slower.
This is because ES-KMeans+ applies dynamic programming iteratively, while the simple approach only segments and clusters once.

\subsection{Effect of our Design Choices}

Our simple approach gives competitive performance by combining good word boundaries with good self-supervised features for clustering (last line, Table~\ref{tbl:zrc_results}).
To support this argument, we look at how performance is impacted when an older boundary detection method is used or when conventional speech features are employed.
The last line of Table~\ref{tbl:ablation} gives the performance of our final system.
The first line is a system using subsampled MFCCs (as in the original ES-KMeans)
instead of self-supervised features.
This comparison shows that modern features are much better for lexicon building:
NED worsens by 34.9\% absolute when using MFCCs.
A comparison
between lines two and three also
shows that the prominence-based approach of~\cite{ankita_tti} performs much better than when the SylSeg boundaries are used. 
The combination of improved boundaries and improved features are therefore what leads to the competitive performance of our simple approach. But both
of these improvements can actually 
be attributed to the representation capabilities of self-supervised speech models: 
SylSeg boundaries are computed directly on the raw speech while the prominence-based approach takes advantage of the dissimilarity of encoded features.

\begin{table}[t]
    \mytable
    \captionsetup{justification=centering}
    \caption{Runtime (min) of prominence segmentation with clustering and ES-KMeans+.}
    \setlength{\tabcolsep}{5.7pt}
    \begin{tabularx}{\linewidth}{@{}l@{}c@{\ \ }c@{\ \ }c@{\ \ }c@{\ \ }c@{}}
        \toprule
        Model & English & French & Mandarin & German & Wolof\\
        \midrule
        ES-KMeans+~[ours] & 765 & 330 & 5 & 316 & 8 \\
        Prom. Seg. Clus.~[ours] & \textbf{146} & \textbf{68} & \textbf{1} & \textbf{50} & \textbf{3} \\
        \bottomrule
    \end{tabularx}
    \label{tbl:runtime}
\end{table}

\begin{table}[t]
    \mytable
    \captionsetup{justification=centering}
    \caption{Ablation scores (\%) for the main components of prominence segmentation with clustering on LibriSpeech dev-clean. A boundary tolerance of 20 ms is~allowed.}
    \begin{tabularx}{\linewidth}{@{}lCCC@{}}
        \toprule
        Components & NED & $R$-value & Token $F_{1}$\\
        \midrule
        MFCC + Prom. Seg. & 75.3 & \textbf{50.7} & \textbf{15.6}\\
        HuBERT + SylSeg & 41.8 & 39.1 & 7.5\\
        HuBERT + Prom. Seg. & \textbf{40.4} & \textbf{50.7} & \textbf{15.6}\\
        \bottomrule
    \end{tabularx}
    \label{tbl:ablation}
\end{table}

\subsection{Impact of the Languages in HuBERT Pre-Training}
\label{ssec:result_pretrain}

\begin{figure}[!t]
    \centerline{\includegraphics[width=\linewidth]{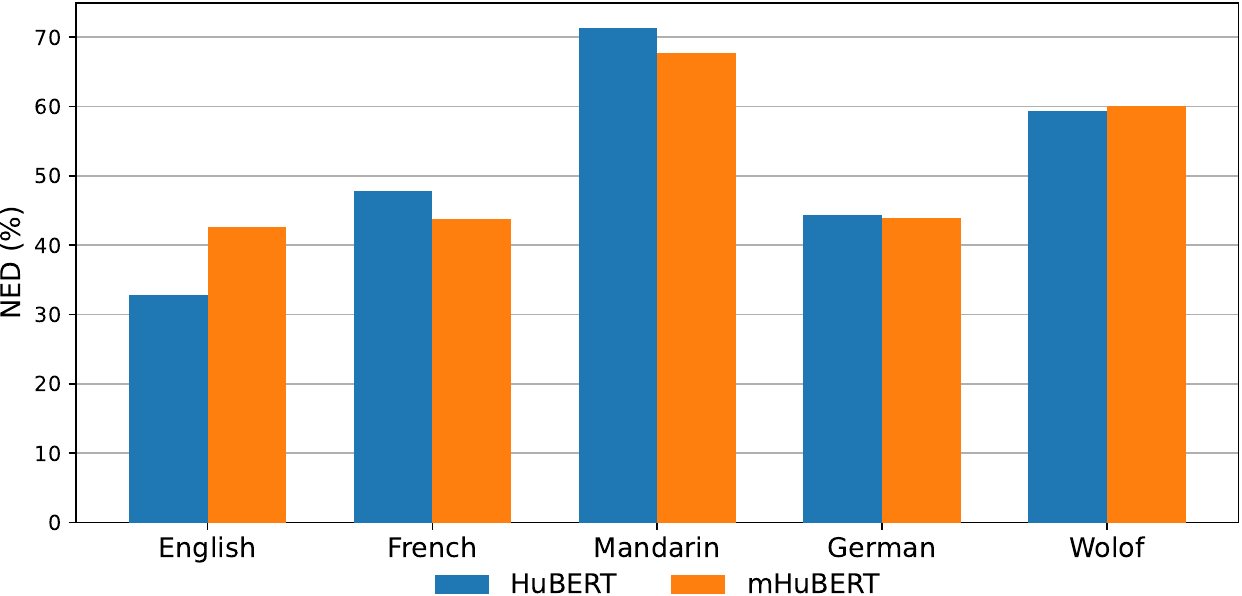}}
    \caption{Normalized edit distance (\%)
    of our prominence-based approach when swapping English HuBERT features for multilingual HuBERT (mHuBERT) features.
    }
    \label{fig:zrc_multilingual_results}
\end{figure}

Up to now, 
all experiments were conducted using the English HuBERT model, even when applying the approach to non-English
data.
To investigate the effect of the feature encoder's pre-training language, we use the recent multilingual HuBERT (mHuBERT) model~\cite{mhubert}. 
Specifically, we inspect the lexicon-building ability of this model compared to the English-specific setup.
We find that the 8th mHuBERT layer performs the best overall and, using the same boundaries as in Table~\ref{tbl:zrc_results}, we cluster the word segments.\footnote{Another option is to use mHuBERT boundaries, but by keeping them
constant, we specifically assess the impact of the features on the lexicon.}
Figure~\ref{fig:zrc_multilingual_results} shows the resulting NED scores. 

For all non-English languages, mHuBERT gives similar or better performance compared to the English model.
For English, the language-specific HuBERT performs 9.8\% absolute better than mHuBERT.
(English is one of the 147 training languages used in mHuBERT.) 
It therefore seems beneficial to train a self-supervised speech model specifically for the language on which it will be applied.
To further investigate this, we perform a test using a Mandarin-specific HuBERT,\footnote{\url{https://huggingface.co/TencentGameMate/chinese-hubert-base}} also extracting features from the 8th transformer layer while keeping the boundaries consistent. 
Here, the NED improves from 67.7\% to 61.9\%.\footnote{Open-source HuBERT models are not available for the other languages.}

Overall, our results show that there may be small benefits from using multilingual training rather than a monolingual model trained on a language different from the target, but language-specific self-supervised training remains the best. 
Similar findings have been made in other recent work~\cite{dusted}.

\section{Conclusion}
\label{sec:conclusion}

This paper showed that unsupervised word segmentation and lexicon learning can be performed competitively by combining a simple boundary detection method with clustering on modern self-supervised features.
Concretely, boundary detection is performed using dissimilarities between adjacent self-supervised features and $K$-means clustering is then performed on averaged features to obtain a lexicon.
This simple method was compared to state-of-the-art dynamic programming methods, including our own updated version of the embedded segmental $K$-means (ES-KMeans) approach.
While our ES-KMeans+ method gave new state-of-the-art results on several metrics in the ZeroSpeech benchmarks, our simple prominence-based segmentation and clustering method performed competitively while being much faster.
We showed that both the initial boundaries and clustering features are important to achieve good performance.
We also showed that even better performance is possible if the self-supervised speech model is trained specifically on data from the particular testing language.

\clearpage
\balance{}
\bibliographystyle{IEEEtran}
\bibliography{IEEEabrv,refs}

\end{document}